\documentclass[journal=ancac3,manuscript=letter,layout=twocolumn]{achemso}
\usepackage[version=3]{mhchem}

\author{Fabian Donat Natterer}
\altaffiliation{Current address: Center for Nanoscale Science and Technology, NIST, Gaithersburg, MD 20899, USA}
\author{Fran\c{c}ois Patthey}
\affiliation{Institute of Condensed Matter Physics (ICMP), \'Ecole Polytechnique F\'{e}d\'{e}rale de Lausanne (EPFL), Station 3, CH-1015 Lausanne, Switzerland}
\author{Harald Brune}
\email{harald.brune@epfl.ch}
\affiliation{Institute of Condensed Matter Physics (ICMP), \'Ecole Polytechnique F\'{e}d\'{e}rale de Lausanne (EPFL), Station 3, CH-1015 Lausanne, Switzerland}

\title{Rotational Excitation Spectroscopy with the STM through Molecular Resonances}
\abbreviations{STM, IETS, STM-RES}

\begin{document}

\begin{abstract}
We investigate the rotational properties of molecular hydrogen and its isotopes physisorbed on the surfaces of graphene and hexagonal boron nitride ($h$-BN), grown on Ni(111), Ru(0001), and Rh(111), using rotational excitation spectroscopy (RES) with the scanning tunneling microscope. The rotational thresholds are in good agreement with $\Delta J=2$ transitions of freely spinning para-\ce{H2} and ortho-\ce{D2} molecules. The line shape variations in RES for \ce{H2} among the different surfaces can be traced back and naturally explained by a resonance mediated tunneling mechanism. RES data for \ce{H2}/$h$-BN/Rh(111) suggests a local intrinsic gating on this surface due to lateral variations in the surface potential. An RES inspection of \ce{H2}, \ce{HD}, and \ce{D2} mixtures finally points to a multi molecule excitation, since either of the three $J=0\rightarrow2$ rotational transitions are simultaneously present, irrespective of where the spectra were recorded in the mixed monolayer.
\end{abstract}

\section{Introduction}
Many of the intriguing properties of molecular hydrogen can be traced back to the interplay of its two alike nuclei. The two nucleons are fermions and therefore the total molecular wave function is anti-symmetric with respect to proton permutation; a consequence of the Pauli principle. Since the vibrational and the electronic $( ^1 \Sigma ^+ _g)$ ground states are symmetric, either the rotational or the nuclear wave functions have to be anti-symmetric. Therefore the antisymmetric nuclear singlet state (para) requires a symmetric rotational wave function (even rotational quantum number $J$), whereas the symmetric nuclear triplet state (ortho) implies an antisymmetric rotational wave function (odd $J$). Without inhomogeneous magnetic fields, the conversion into the other nuclear spin isomer is forbidden by symmetry. Therefore each of the two nuclear spin isomers remains in its rotational subspace of either solely even or merely odd $J$ states. An interesting consequence from this strict correlation is that the knowledge of the rotational state provides information about the nuclear spin state, and vice versa. By way of example, the rotational energy of such a diatomic rigid rotor in terms of $J$ and $B$ is $E = BJ(J+1)$, where $B$ is the rotational constant. Note that similar thoughts likewise apply to deuterium, and in more general terms, they apply to any homonuclear diatomic and polyatomic molecule; some representative examples thereof with their $B$ values are listed in table~\ref{table1}. The existence of nuclear disparate forms of H$_2$ with room temperature equilibrium abundance of 1:3 for para:ortho, has been predicted by Heisenberg~\cite{hei27}. The first preparation of pure para-\ce{H2} enabled the confirmation of this prediction by emission spectra and heat conductivity measurements in the gas phase~\cite{bon29}.

The scanning tunneling microscope (STM) allows to harvest the great potential of inelastic electron tunneling spectroscopy (IETS)~\cite{lam68,wol12} for the study of vibrational~\cite{sti98a,lau99,sti98b} and magnetic~\cite{hei04,ott08,kha11,don13} properties of individual atoms and molecules. It was not until recently that the the capabilities of IETS were extended to study the rotational excitations of physisorbed hydrogen~\cite{nat13a,li13}. These contributions demonstrated the great potential of rotational excitation spectroscopy (STM-RES) as a tool for characterizing bond lengths, chemical identity, and, notably the molecular nuclear spin states. The two major obstacles that prevented STM-RES from advancing were the small rotational constants, $B=\hbar^2/2I$ where $I$ is the molecule's moment of inertia, and the strong interaction of molecules with most metal surfaces. The former limitation can be lifted by employing low temperatures and small modulation amplitudes. The latter constraint can be overcome by studying physisorbed molecules on ultra-thin decoupling layers, since the molecules are otherwise frequently stuck in a hindered or frustrated rotational motion~\cite{lau99,sti98b}. However, only a better knowledge of the actual excitation mechanism in STM-RES will allow for a more directed research what would ultimately enable the examination of molecules with even smaller rotational constants.

\begin{table}
\caption{\small Rotational constants $B=\hbar^2/2I$ for selected diatomic and polyatomic molecules with indistinguishable nuclei in their vibrational and electronic ground state~\cite{web}. Polyatomic molecules have rotational constants for different symmetry axes.}
\label{table1}
\vspace{0.0cm}
\centering
\begin{tabular}{ll}
\hline\hline
Molecule        & $B$ (meV) \\ \hline
H$_2$ & 7.36 \\
HD & 5.54  \\
D$_2$ & 3.71  \\
$^{14}$N$_2$ & 0.247  \\
$^{16}$O$_2$ & 0.178 \\\hline
CH$_4$& 0.65 / 0.65 / 0.65\\
H$_2$O & 3.46 / 1.80 / 1.15 \\
NH$_3$ & 1.17 / 1.17 / 0.77 \\\hline \hline
\end{tabular}
\end{table}

The interaction of H$_2$ with solid surfaces is either strong, where it leads to the dissociation and the subsequent chemisorption of single H atoms, or it is weak, where the molecules are merely bound by van der Waals (vdW) forces in a physisorbed state that is stable at low temperature. The latter case is of interest for the detection of the two nuclear spin isomers for surface adsorbed H$_2$. This has been achieved with neutron diffraction~\cite{nie77, fra88}, nuclear magnetic resonance~\cite{kub85, kim97}, and high-resolution electron energy loss spectroscopy (HREELS)~\cite{and82, avo82, pal87}. HREELS detected the $J = 0 \rightarrow 2$ transition of para-\ce{H2} by an electron energy loss of 46 -- 49~meV, a very weak loss peak at 70~meV has been attributed to the $J = 1 \rightarrow 3$ transition of ortho-\ce{H2}~\cite{avo82}. The respective gas phase values are 43.9 and 72.8~meV~\cite{sil80, sou86}. The weak ortho-\ce{H2} signal decayed after a few minutes~\cite{avo82}, which has been interpreted as ortho to para-\ce{H2} conversion by short range magnetic interactions with the surface~\cite{ili92}. Evidence for rotational excitations has also been seen for the lowest energy $J = 0 \rightarrow 1$ of HD by high-resolution inelastic He atom scattering~\cite{tra04}.

\begin{figure}[t]
\begin{center}
\includegraphics[width = 8.5 cm]{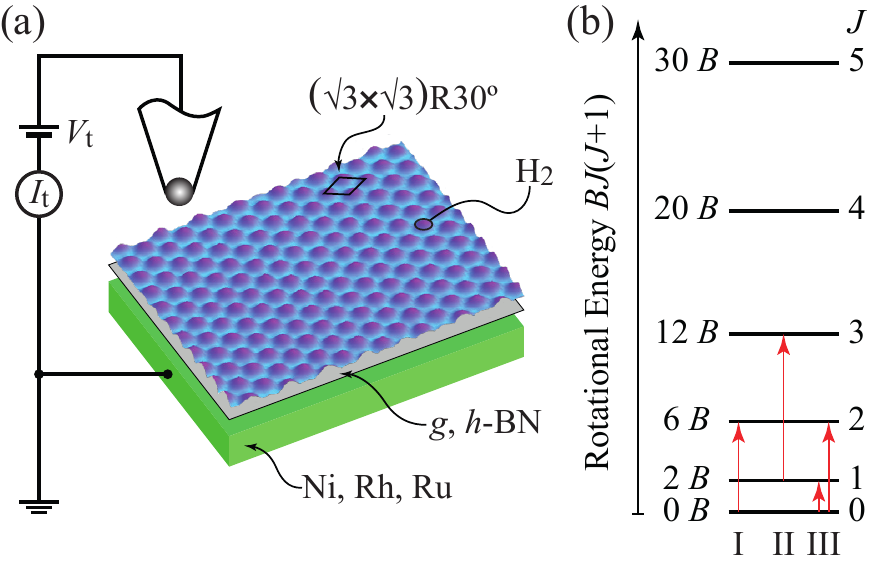}
\end{center}
\caption{\footnotesize  (a) Schematic of experimental setup showing the hydrogen monolayer, physisorbed on either graphene ($g$) or hexagonal boron nitride ($h$-BN), grown on a metallic substrate (Ni, Rh, or Ru). The hydrogen $(\sqrt3\times\sqrt3)\mathrm{R}30^{\circ}$ superstructure is indicated by the lozenge. (b) Rotational level diagram for diatomic rigid rotor, expressed in units of the rotational constant $B=\hbar^2/2I$, where $I$ denotes the moment of inertia. The diagram is drawn for a population of (I) solely even (para-\ce{H2}, ortho-\ce{D2}, ortho-$^{14}$\ce{N2} ...), (II) uniquely odd (ortho-\ce{H2}, para-\ce{D2}, para-$^{14}$\ce{N2}, $^{16}$\ce{O2},..), and (III) both rotational levels.}
\label{setup}
\end{figure}

We recently demonstrated STM-RES for H$_2$, HD, and D$_2$, physisorbed on hexagonal boron nitride ($h$-BN)~\cite{nat13a}. In parallel, Li~\textit{et. al.}~\cite{li13} published RES data for hydrogen and its isotopes on Au(110) that displayed small inelastic conductance steps at the energies expected for RES. Both studies revealed that the molecules shared the properties of their gas phase counterparts and correspondingly showed the respective transitions $J=0\rightarrow2$ with similar rotational constant. These contributions demonstrated the nuclear spin sensitivity of STM-RES and helped to identify the nuclear distinct para-\ce{H2} and ortho-\ce{D2} spin isomers with unmatched spatial selectivity~\cite{nat13a,li13}. Furthermore, the transitions $J=0\rightarrow 2$ and $J=0\rightarrow 1$ were observed for HD, as expected for diatomics with distinguishable nuclei~\cite{nat13a}.


These two previous studies motivated the present investigation of STM-RES, in particular in view of establishing a better understanding of the actual excitation mechanism. To this end, we measured STM-RES for hydrogen and deuterium on four substrates. These were graphene ($g$) and hexagonal boron nitride ($h$-BN) grown on Ru(0001), Ni(111), and Rh(111), as illustrated in Fig.~\ref{setup}. We could reconcile the subtle variations in the spectroscopic signatures in STM-RES by a resonance mediated tunneling mechanism and can now also rationalize why former STM-RES signatures were fully quenched~\cite{gup05,tem08, wei10} or strongly suppressed~\cite{li13} by a metal substrate. Furthermore, this model exposes how one could control the molecular dynamics by externally accessible tuning nobs. Finally, experiments with isotopic mixtures of H$_2$, HD, and D$_2$ strengthen the multi-molecule picture of STM-RES for these molecules.


\section{Results and discussion}
The four investigated substrate combinations are structurally equivalent and only differ by little in their lattice parameters. They also exhibit similar behavior when grown on the various metal substrates. The lattice mismatch and the tendency of carbon and nitrogen to strongly interact with the transition metal substrate (except for coinage metals) leads to a distinct inhomogeneous bonding landscape of periodically interrupted regions of strong hybridization and mere dispersive vdW interaction. For the case of $g$/Ru, the resulting topography is an array of nanometer sized vdW bound hills with a period of 3~nm~\cite{mar07}, surrounded by strongly hybridized valleys, in the notorious $g$/Ru(0001)--$(23\times23)$ structure~\cite{mar08,ian13a}. In comparison, $h$-BN adopts the inverted topography on Rh(111) and a lower areal fraction is strongly bound as a consequence of the two unlike atoms in $h$-BN. The hybridization of the N lone pair orbital with the Rh $d$-states~\cite{din11} characterizes the array of depressions with a period of 3.22~nm~\cite{bun07} in the $h$-BN/Rh(111)--$(12\times12)$ structure~\cite{cor04,las07}. For the lattice matched systems of $g$/Ni(111) and $h$-BN/Ni(111) the layers grow almost perfectly flat and in the commensurate $(1\times1)$ structure~\cite{gam97,auw99}. The $h$-BN and $g$ layers are strongly bound on this lattice matched substrate~\cite{auw99,gra03,gam97,pre04}.

\begin{figure*}[t!]
\begin{center}
\includegraphics[width = 17.8 cm]{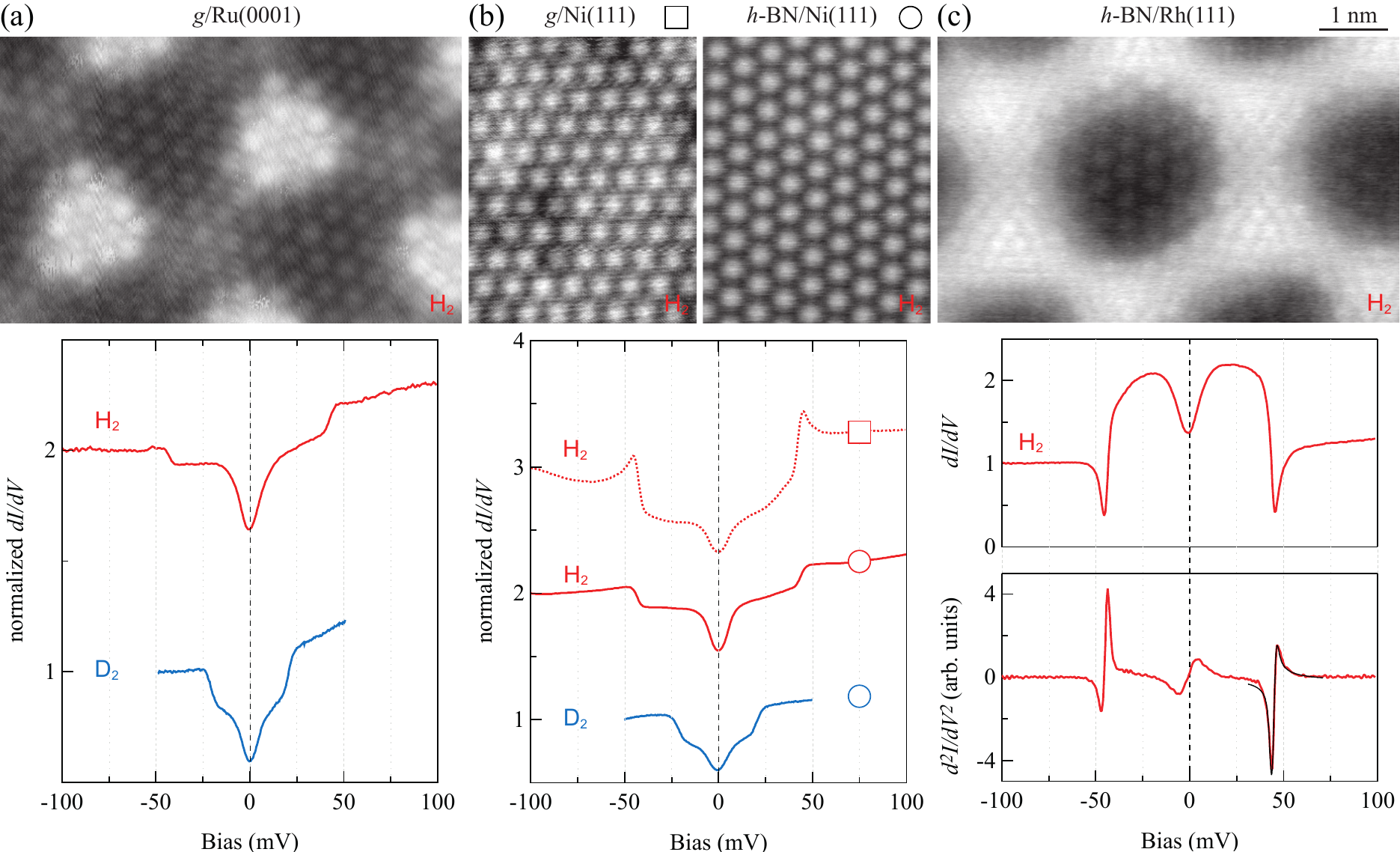}
\end{center}
\caption{\footnotesize  STM images of hydrogen superstructures on graphene and hexagonal boron nitride (top) with the corresponding rotational excitation spectra (bottom). (a) The hydrogen molecules adsorb in the $(\sqrt3\times\sqrt3)\mathrm{R}30^{\circ}$ superstructure in the valleys of the $g$/Ru(0001)--$(23 \times 23)$ moir{\'e} unit cell. The $dI/dV$ traces were recorded in their center and the curves show averages of 201 (H$_2$) and 56 (D$_2$) point spectra. (b) Hydrogen superstructure on $g$/Ni(111)--$(1\times1)$ to the left, and on $h$-BN/Ni(111)--$(1\times1)$ to the right. The $dI/dV$ data represent averages of 156 (H$_2$/$g$), 889 (H$_2$/$h$-BN), and 193 (D$_2$/$h$-BN) spectra. (c) Around 20 hydrogen molecules are trapped in the depressions of the $h$-BN/Rh(111)--$(12 \times 12)$ unit cell. The $dI/dV$ curve was recorded in the center of the moir\'e depressions and represents an average of 289 single spectra. The superimposed fit with Eq.~(\ref{con}) (black) shows excellent agreement with the measured $d^2I/dV^2$ spectra. [$T_{\rm STM} = 4.7$~K, (a) $V_{\rm t} = -20$~mV, $I_{\rm t} = 50$~pA, (b--left) $V_{\rm t} = -50$~mV, $I_{\rm t} = 50$~pA, (b--right) $V_{\rm t} = -20$~mV, $I_{\rm t} = 20$~pA, (c) $V_{\rm t} = -20$~mV, $I_{\rm t} = 5$~pA.].}
\label{STM-didv}
\end{figure*}

Although the surface topographies appear quite distinct, the common trait of all four systems are the regions where $g$ and $h$-BN are strongly bound. This connecting feature is indeed also reflected in the preferred regions for the condensation in the well known $(\sqrt3\times\sqrt3)\mathrm{R}30^{\circ}$ superstructure, as seen in the STM images of Fig.~\ref{STM-didv}. This superstructure has previously been reported for the surfaces of graphite~\cite{nie77,seg82,kub85} and of $h$-BN~\cite{kim99} bulk samples, and recently on monolayer $h$-BN/Ni(111)~\cite{nat13a}. The molecules reside on the six-fold coordinated hollow site in the center of the C$_6$ or (BN)$_3$ rings. On the strongly bound stacking areas of the respective moir\'e pattern, these sites are aligned with the threefold hollows sites of the underlying metal substrate.

We recorded the differential conductance ($dI/dV$) of physisorbed hydrogen and deuterium on the four surfaces in the regions of the $(\sqrt3\times\sqrt3)\mathrm{R}30^{\circ}$ superstructure, as is indicated in the bottom part of Fig.~\ref{STM-didv}. Irrespective of the underlying substrate, the measured $dI/dV$ spectra show pairs of sudden conductance steps at the threshold energies of $\pm44$~meV and $\pm21$~meV for hydrogen and deuterium, respectively. Note that the low-energy excitations ($\le5$~meV) were ascribed to phonon-excitations of the molecular layers~\cite{fra88, lau89, jan91, nat13a} and are not discussed in the following. The regarded $dI/dV$ signatures are characteristic of the rotational transitions $J=0\rightarrow2$ of hydrogen and deuterium with energies of $6B$ [cf. Fig.~\ref{setup}~(b)]. These values are in good agreement with previous STM-RES reports for hydrogen on $h$-BN/Ni(111)~\cite{nat13a} and on Au(110)~\cite{li13}. The close coincidence of the measured rotational energies for physisorbed hydrogen with the gas phase values serves as an indicator of the weak surface-molecule interaction and illustrates how the decoupling layers $g$ and $h$-BN can be used to study intrinsic molecular properties with STM-RES. One notable information that can be extracted is the distinction of the nuclear spin isomers para-\ce{H2} and ortho-\ce{D2} due to their characteristic rotational energies of $6B$; the other isomer would have $10B$\footnote{The absence of the complementary isomers is attributed to a fast conversion  of the $J = 1$ to the $J = 0$ nuclear spin isomer of the respective isotope~\cite{ili92,ili13}}.

A closer analysis of the line shape of the spectra in Fig.~\ref{STM-didv} reveals valuable information about the excitation mechanism as is shown in the following. While we observe an increasing differential conductance at the excitation threshold for hydrogen on $g$/Ru(0001), $g$/Ni(111), and $h$-BN/Ni(111), a negative differential resistance (NDR)~\cite{bed89,lyo89} is seen for hydrogen on $h$-BN/Rh(111) which in turn shares resemblance with an asymmetric Fano-line shape~\cite{fan61}. Earlier STM contributions described the NDR for tunneling across adsorbed molecules as a result of a conformational change of the molecule in the junction~\cite{gau00}, or described the NDR in terms of a two-state switching between two levels with different conductance~\cite{gup05}.

In the following, however, we will interpret our observations by a resonance mediated tunneling process which naturally explains both, NDR and stepwise increase in differential conductance at the excitation threshold, as well as all the intermediate cases [such as for H$_2$/$g$/Ni(111) in Fig.~\ref{STM-didv}~(b)]. We employ the formalism of Persson and Baratoff~\cite{per87, bar88} which was initially devised for inelastic electron tunneling via molecular resonances. Although the model was originally developed for the excitation of vibrational modes for chemisorbed molecules, it is safe to also include molecular rotations for a physisorbed molecule in its description provided that the following analogy applies. The model assumes a broad and short lived resonance that is formed due to a strong hybridization of the molecule with the surface. However, a broad resonance state can also exist in form of a negative ion resonance (also referred to as shape resonance)~\cite{sul73,pal92} for molecules that are only weakly perturbed by the surface. The incoming electron can then either be temporarily trapped in this  broad resonance state of the molecule and relax via all available degrees of freedom, for instance by inelastically transferring a vibrational or rotational quantum to the molecule, or the electron can tunnel elastically into the continuum instead. This fundamental relationship is encountered for a wide range of physical processes and emerges from the quantum interference between a discrete set of states with a continuum.

\begin{figure}[t!]
\begin{center}
\includegraphics[width = 8.5 cm]{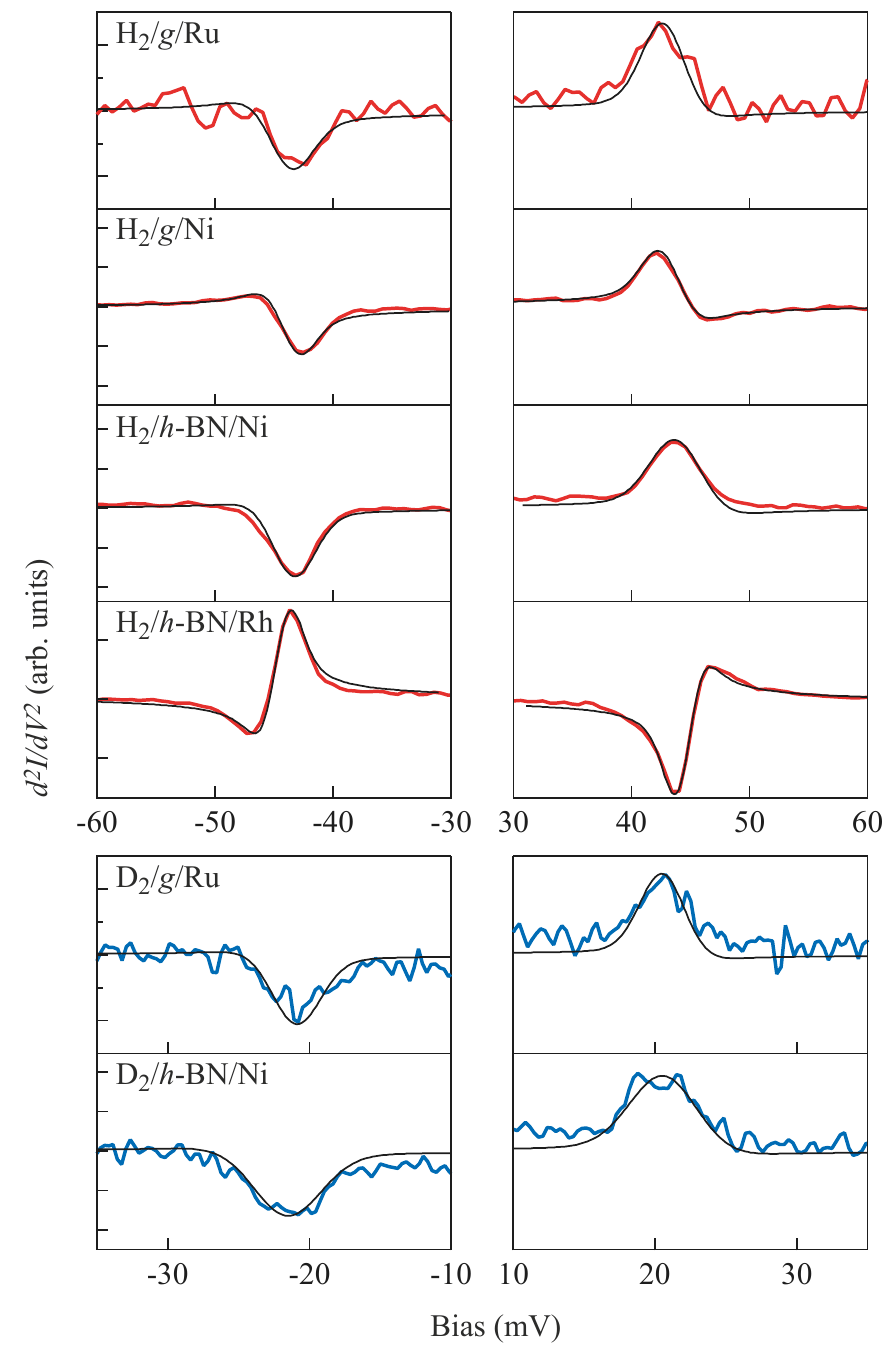}
\end{center}
\caption{\footnotesize Experimental $d^2I/dV^2$ data (red, blue) around the $J=0\rightarrow2$ rotational thresholds with superimposed simulated spectra (black) using Eq.~\ref{con} and the fit parameters of table~\ref{fit-param}.}
\label{fits}
\end{figure}

The model of Persson and Baratoff relies on the position of the molecular resonance ($\epsilon _a$) with respect to the Fermi energy, the resonance width ($\Gamma$), the excitation threshold ($\Omega$), and a coupling parameter ($\delta\epsilon$). The differential conductance ($\sigma$) as a function of the applied voltage ($V_{\rm t}$) and in proximity of the threshold is described as:
\begin{align}
\label{con}
\sigma&=\frac{\delta \epsilon ^2}{(\epsilon _a+ {\rm e}V_{\rm t})^2+(\Gamma/2)^2} \nonumber \\
 & \times \left\{ \frac{(\epsilon _a +{\rm e}V_{\rm t} -\Omega)^2-(\Gamma /2)^2}{(\epsilon _a +{\rm e}V_{\rm t}-\Omega)^2+(\Gamma /2)^2}\Theta({\rm e}V_{\rm t}-\Omega) \right. \nonumber \\
& \left. -\frac{\Gamma}{\pi}\frac{\epsilon _a +{\rm e}V_{\rm t} - \Omega}{(\epsilon _a +{\rm e}V_{\rm t} -\Omega)^2+(\Gamma /2)^2} \ln \left| \frac{{\rm e}V_{\rm t}-\Omega}{\Delta}\right| \right\}.
\end{align}
From this relationship, we can appreciate the natural appearance of NDR whenever $\left|\epsilon _a +{\rm e}V_{\rm t} -\Omega\right|<\Gamma /2$~\cite{bar88}. The character of the observed features is therefore a sensitive gauge for the alignment of the molecular resonance with respect to the Fermi energy.

\begin{table*}[t!]
\caption{\footnotesize Fit parameters for simulated $d^2I/dV^2$ spectra in Fig.~\ref{STM-didv}~(c) and Fig.~\ref{fits}, using Eq.~(\ref{con}). The symbols $\Omega$, $\epsilon _a$, and $\Gamma$ are the rotational excitation threshold, the position of the molecular resonance with respect to the Fermi energy, and the width of the molecular resonance, respectively. The errors were determined by varying the parameters until the disagreement with experimental spectra became significant.}
\label{fit-param}
\vspace{0.0cm}
\centering
\begin{tabular}{lcccc}
\hline\hline
                    & $\Omega$ (meV)  & $\epsilon _a$ (meV) & $\Gamma$ (meV) & $T_{\rm E}$ (K)\\ \hline
\ce{D2}/$g$/Ru      & $20.8\pm0.1$  & $660\pm40$    & $100\pm60$& $7.7\pm0.5$              \\
\ce{H2}/$g$/Ru      & $42.8\pm0.2$  & $660\pm40$    & $120\pm50$& $8.7\pm0.5$              \\
\ce{H2}/$g$/Ni      & $43.0\pm0.5$  & $750\pm30$    & $500\pm60$& $7.7\pm0.5$               \\
\ce{H2}/$h$-BN/Ni   & $44.1\pm0.2$  & $1200\pm20$   & $310\pm90$& $10.7\pm0.5$              \\
\ce{D2}/$h$-BN/Ni   & $20.8\pm0.1$  & $1200\pm80$   & $170\pm100$&$9.7\pm0.5$               \\
\ce{H2}/$h$-BN/Rh   & $44.5\pm0.1$  & $-200\pm20$   & $820\pm40$& $5.0\pm0.5$              \\ \hline
\end{tabular}
\end{table*}

We approximated the measured spectra by iteratively adjusting the above parameters and by taking thermal and modulation broadening into account~\cite{lam68, kle73}. We achieved excellent agreement between simulated and experimental $d^2I/dV^2$ spectra with the parameters found in table~\ref{fit-param}, and exemplarily displayed for H$_2$/$h$-BN/Rh(111) in Fig.~\ref{STM-didv}~(c), and in more detail shown around the rotational threshold for all substrates in Fig.~\ref{fits}. The NDR found for H$_2$/$h$-BN/Rh(111) can now be rationalized through a shift of the resonance to $\epsilon _a=-200$~meV, whereas the resonance position of the other systems is found around $+1$~eV. The $h$-BN/Rh(111) system is particular, in that the $(12\times12)$ moir\'e pattern is accompanied by a periodic modulation of the work function between the depressions and the wire structure. These work function modulations lead to electrostatic dipole rings~\cite{dil08} that were shown to immobilize molecules and atoms~\cite{cor04,dil08,nat12a}. We can therefore attribute the shift of the molecular resonance position for H$_2$/$h$-BN/Rh(111) to the local variations of the electrostatic potential on this surface. Consequently, this system is a prototype for the effect of molecular gating and presents an interesting perspective for further STM-RES studies. Note that the interplay of the resonance position and the resonance width may lead to a suppression of the rotational excitation. The former could be shifted, for instance, by tuning the sample work function and would thereby provide a powerful control of the molecular dynamics. It appears that an unfortunate alignment or improper width of the resonance may be the reason why earlier contributions missed to report STM-RES for hydrogen on copper and silver~\cite{gup05,tem08}, and why the intensity of the rotational excitations was small for hydrogen on Au(110)~\cite{li13}. The interaction with the metal substrate can furthermore lead to an additional broadening of the molecular resonance and thereby reduce its lifetime with the consequence that an energy transfer from the electron to the inelastic channel becomes inefficient\footnote{as was pointed out by William Gadzuk, NIST}. Furthermore, it is curious to note that the Smoluchowski effect and the involved local work-function variations for the missing row reconstructed Au(110) surface could result in the fortunate shifting of the molecular resonance~\cite{smo41}.

When we take another look at our experimental data we can extract additional information from the simulations with Eq.~(\ref{con}) and their approximation to our $d^2I/dV^2$ spectra. Firstly, the RES thresholds show some variation that can be attributed to subtle compressions of $-0.7$\% and extensions of $+1.3$\% in the bond length with respect to the free molecule~\cite{li13}. Secondly, the intrinsic width of the excitation ($W_{\rm I}$, half-width-at-half maximum) is related to the lifetime ($\tau$) of the rotationally excited state by $\tau = \hbar/W_{\rm I}$. The widths were determined by introducing an effective temperature in the broadening kernel of the above mentioned convolution (cf. table~\ref{fit-param}). The extracted mean lifetimes range between 10~ps and 500~fs. Calculations reported weakly dispersing roton bands~\cite{jan91} that would broaden the excitation energy implying that the above stated lifetime range represents a lower limit. An upper limit is found when recording $dI/dV$ spectra with gradually increasing current. At 500~pA, one electron tunnels every 320~ps, yet we saw no clear sign of pumping into higher lying rotational quantum states. Therefore the electrons are always probing the molecules in their rotational ground state and the lifetimes of the excited $J = 2$ state are $\tau \ll 320$~ps.


\begin{figure}[t!]
\begin{center}
\includegraphics[width = 8.5 cm]{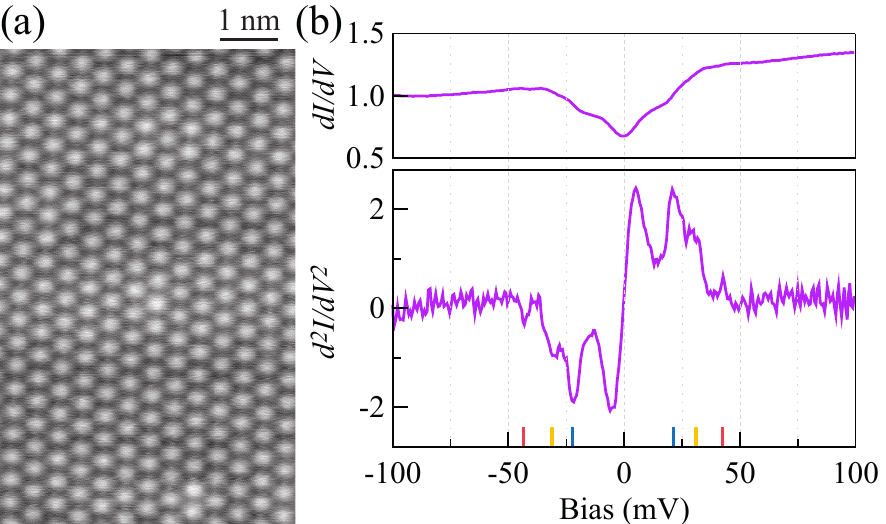}
\end{center}
\caption{\footnotesize Ternary mixture of hydrogen, deuterium, and hydrogen-deuteride. (a) STM image of the mixture showing the $(\sqrt3 \times \sqrt3){\rm R}30^{\circ}$ superstructure and no apparent height contrast between the various molecules. (b) The differential conductance spectrum (top) and its numerical derivative show rotational signatures of all three molecules at around 22, 31, and 43~meV ($V_{\rm t} = -50$~mV, $I_{\rm t} = 50$~pA, average of 1100 spectra $R_{\rm G} = 1$~G$\Omega$).}
\label{mixture}
\end{figure}

Typical differential conductance spectra characterize a selected molecule immediately underneath the tip apex. However, as we showed for the case of hydrogen in STM-RES, the probed ensemble size was considerably larger and amounted to $(60\pm30)$ molecules~\cite{nat13a}. We prepared hydrogen isotope mixtures and found further support for the multi-molecule picture of STM-RES. A heat treated mixture with the equilibrium concentration of H$_2$, HD, and D$_2$ was adsorbed on $h$-BN/Ni(111) at 10~K and subsequently characterized by STM-RES as before. Figure~\ref{mixture}~(a) is devoid of any topographic signs for different molecules and one might doubt the presence of the various hydrogen isotopes. However, the inspection of the $dI/dV$ spectra in Fig.~\ref{mixture}~(b) clearly shows the signatures of all three isotopes with RES at 22, 31, and 43~meV, corresponding to the known $J=0\rightarrow 2$ transitions of deuterium, deuterium-hydride, and hydrogen, respectively. Not only were all three isotopes present, the spectra did notably not reveal what molecule was residing underneath the tip-apex; all three RES were simultaneously present, irrespective of where the spectra had been measured. Mixtures with a commensurate $\sqrt3$ coverage $\Theta\le1$ were shown to equally adopt the $(\sqrt3\times\sqrt3)\mathrm{R}30^{\circ}$ phase and the isotopic molecules are expected to be randomly distributed~\cite{bie99}. This situation is in good agreement with the multi-molecule picture of STM-RES for hydrogen on the regarded surfaces. The presented approach could be used to further study the lateral extent of STM-RES by varying the relative concentration of the mixtures. Note, however, that an enrichment with the heavier molecules can occur for prolonged exposures to the thermodynamic equilibrium isotope mixture through inter-layer transport~\cite{bie99}. The latter could naturally explain the observed intensity variations for H$_2$, HD, and D$_2$ in the spectra of Fig.~\ref{mixture}~(b). Note, however, that the above illustrated scenario is still a tentative explanation. Alternatively, the molecules could be highly mobile, even in the $\sqrt3$ structure. The displayed topography and spectra of Fig.~\ref{mixture} would then merely correspond to temporal averages of the various molecules with different residence times beneath the tip apex. At this point, we cannot decide against or in favor of the latter explanation.

In conclusion, we have shown that STM-RES can be used on a large selection of different substrates. We were able to implement a resonance mediated tunneling model that explains all spectroscopic signatures of hydrogen and of its isotopes. It was found that a local variation in the surface potential can lead to a shift of the molecular resonance position with respect to the Fermi energy. Since these variations can, in principle, be controlled by external means, the latter example serves as a blueprint for controlling the molecular dynamics of physisorbed molecules. Exploring the resonance mediated tunneling model will furthermore ease the study of alternative molecules that have less pronounced RES transitions, with good starting points proposed in table~\ref{table1}.
\section{Methods}
{\small The measurements were performed with a homemade low temperature STM, operating at 4.7~K and in an ultra-high vacuum (UHV) chamber with a base pressure $p_{\rm tot} < 5 \times 10^{-11}$~mbar~\cite{gai92}. We prepared atomically clean single crystals of Rh(111), Ni(111), and Ru(0001) by repeated cycles of Ar$^+$ sputtering (10~$\mu {\rm A/cm^2}$, 1~kV, 300~K, 30~min), annealing in oxygen (815~K, 5~min, $2 \times 10^{-7}$~mbar), and flash to 1450, 1100, and 1400~K, respectively. Monolayers of hexagonal boron nitride were grown by chemical vapor deposition (CVD) on Ni~\cite{nag95,pre04,nat13b}, and on Rh~\cite{cor04,pre07,nat12b} at 1040~K with a borazine partial pressure of $2 \times 10^{-6}$~mbar and 3 minutes exposure. Graphene was either grown by CVD on Ni~\cite{gam97} at 800~K or on Ru~\cite{log09,nat12a} at 1000~K, using an ethylene partial pressure of $1 \times 10^{-6}$~mbar for 10~min, or it was prepared by simply heating a Ru crystal to 1200~K that had experienced several CVD cycles before~\cite{sut08,nat12a}. Subsequent to the layer growth, the samples were cooled to 10~K and exposed to hydrogen. We dosed molecular hydrogen, deuterium, and a mixture of both, by backfilling the UHV chamber through a leak-valve. The H$_2$/D$_2$ mixture was heated to 500~K in a dedicated volume in order to accelerate the transition into the thermodynamic equilibrium with a well defined H$_2$:HD:D$_2$ ratio of 1:2:1. Note that all pressure values state the readout of a N$_2$--calibrated ionization gauge. The STM images were acquired at constant current, and the indicated tunnel voltages $(V_{\rm t})$ correspond to the sample potential (cf. Fig~\ref{setup}). Scanning tunneling spectroscopy was performed by recording the bias-dependent differential conductance ($dI/dV$) using a lock-in amplifier and adding a sinusoidal 2~mV peak-to-peak modulation at 397~Hz to the bias voltage. For every point spectrum, the tip was first stabilized at negative bias with $R_{\rm G} = 1$~G$\Omega$, the feedback-loop was then opened, and a single $dI/dV$ curve was recorded within 40~s. The spectra were offset by one unit for clarity. $d^2I/dV^2$ curves were obtained by numerical derivation of $dI/dV$ data. Both, tungsten and platinum-iridium tips were used for spectroscopy.}

\begin{acknowledgement}
{\small We gratefully acknowledge funding from the Swiss National Science Foundation (SNSF). FDN also acknowledges financial support from SNSF Early Postdoc.Mobility fellowship under project number 148891.}
\end{acknowledgement}
{
\providecommand{\latin}[1]{#1}
\providecommand*\mcitethebibliography{\thebibliography}
\csname @ifundefined\endcsname{endmcitethebibliography}
  {\let\endmcitethebibliography\endthebibliography}{}

}
\end{document}